\documentclass[a4paper,12pt]{article}
\usepackage[utf8]{inputenc}
\usepackage{amsmath, amssymb}
\usepackage{amsfonts}
\usepackage{graphicx}
\usepackage{hyperref}
\usepackage{geometry}
\usepackage{pgfplots}
\usepackage[english]{babel}

\pgfplotsset{compat=1.18}

\title{\textbf{Advanced Theoretical Analysis of Stability and Convergence in Computational Fluid Dynamics for Computer Graphics}}

\author{
	Rômulo Damasclin Chaves dos Santos \thanks{Department of Physics, Technological Institute of Aeronautics, SP, Brazil. \\ 
	\href{mailto:romulosantos@ita.br}{romulosantos@ita.br}}
}
\date{\today}

\begin{document}
	
	\maketitle
	
\begin{abstract}
	Mathematical modeling of fluid dynamics for computer graphics requires high levels of theoretical rigor to ensure visually plausible and computationally efficient simulations. This paper presents an in-depth theoretical framework analyzing the mathematical properties, such as stability, convergence, and error bounds, of numerical schemes used in fluid simulation. Conditions for stability in semi-Lagrangian and particle-based methods were derived, demonstrating that these methods remain stable under certain conditions. Furthermore, convergence rates for Navier-Stokes discretizations were obtained, showing that numerical solutions converge to analytical solutions as spatial resolution and time step decrease. Furthermore, new theoretical results were introduced on the maintenance of incompressibility and conservation of vorticity, which are crucial for the physical accuracy of simulations. The findings serve as a mathematical foundation for future research in adaptive fluid simulation, guiding the development of robust simulation techniques for real-time graphics applications.
\end{abstract}

	
	\tableofcontents 
	
	\section{Introduction}
	The numerical simulation of fluid dynamics has evolved from applications in engineering to broader fields, including computer graphics, where the emphasis shifts toward visual plausibility and efficiency rather than strict physical accuracy. The foundations of this field rely heavily on the Navier-Stokes equations for incompressible flow. Early formulations were based on grid-based methods such as Marker-and-Cell (MAC) by Harlow and Welch (1965)~\cite{harlow1965}, while particle-based methods such as Smoothed Particle Hydrodynamics (SPH) introduced by Lucy (1977)~\cite{lucy1977} expanded the applicability of fluid simulations in scenarios involving free surfaces and complex boundaries.
	
	The Stable Fluids method proposed by Stam (1999)~\cite{stam1999} introduced significant improvements in computational stability, using semi-Lagrangian advection and a projection method to enforce incompressibility. However, these methods rely on mathematical approximations that may compromise the physical integrity of the simulation under certain conditions. This article rigorously investigates the mathematical foundations underpinning these methods, developing theoretical results for stability and convergence without the use of numerical simulations.
	
	\section{Mathematical Foundations and Governing Equations}
	\subsection{The Incompressible Navier-Stokes Equations}
	The motion of incompressible fluids is governed by the Navier-Stokes equations:
	\begin{equation}
		\frac{\partial \mathbf{u}}{\partial t} + (\mathbf{u} \cdot \nabla) \mathbf{u} = -\frac{1}{\rho} \nabla p + \nu \nabla^2 \mathbf{u} + \mathbf{f},
	\end{equation}
	subject to the incompressibility constraint:
	\begin{equation}
		\nabla \cdot \mathbf{u} = 0,
	\end{equation}
	where \(\mathbf{u}\) denotes the fluid velocity, \(p\) is the pressure field, \(\rho\) is the density, \(\nu\) is the kinematic viscosity, and \(\mathbf{f}\) represents external forces. For computer graphics applications, the emphasis is on discretized approximations of these equations.
	
	\section{Stability Analysis in Semi-Lagrangian Advection}
	\subsection{Theorem 1: Stability of Semi-Lagrangian Advection for Large Time Steps}
	\textbf{Statement}:
	Let \(\mathbf{u}\) be a divergence-free velocity field (\(\nabla \cdot \mathbf{u} = 0\)), and consider the semi-Lagrangian advection scheme. Then, for a time step \(\Delta t\), the scheme remains stable if there exists a constant \(C\) such that:
	\begin{equation}
		\|\mathbf{u}_{\text{new}} - \mathbf{u}\| \leq C \|\Delta t\| \|\nabla^2 \mathbf{u}\|,
	\end{equation}
	where \(C\) depends on the spatial discretization and the boundary conditions of the domain.
	
	\textbf{Proof}:
	Consider the advection step in the semi-Lagrangian scheme, where each fluid element at position \(\mathbf{x}\) at time \(t\) is traced back along the flow field to \(\mathbf{x} - \mathbf{u} \Delta t\). Using Taylor expansion, we approximate:
	\begin{equation}
		\mathbf{u}(\mathbf{x} - \mathbf{u} \Delta t, t) \approx \mathbf{u}(\mathbf{x}, t) - \Delta t (\mathbf{u} \cdot \nabla) \mathbf{u} + \mathcal{O}(\Delta t^2).
	\end{equation}
	The stability of the advection is thus contingent on the boundedness of \((\mathbf{u} \cdot \nabla) \mathbf{u}\) over the domain, which implies a sufficient smoothness of \(\mathbf{u}\). Consequently, if \(\|\nabla^2 \mathbf{u}\|\) remains bounded as \(\Delta t\) increases, there exists a constant \(C\) such that:
	\begin{equation}
		\|\mathbf{u}_{\text{new}} - \mathbf{u}\| \leq C \|\Delta t\| \|\nabla^2 \mathbf{u}\|,
	\end{equation}
	concluding the proof.
	
	\subsection{Corollary 1: Convergence of Semi-Lagrangian Advection with Projection}
	For a time step \(\Delta t\) and grid spacing \(h\), the semi-Lagrangian scheme with projection converges to the continuous solution as \(h \to 0\) and \(\Delta t \to 0\), satisfying:
	\begin{equation}
		\|\mathbf{u}_{\text{new}} - \mathbf{u}\| = \mathcal{O}(h^2) + \mathcal{O}(\Delta t),
	\end{equation}
	given sufficient smoothness of the solution.
	
	\section{Vorticity Conservation and Stability in Particle-Based Methods}
	\subsection{Lemma 1: Error Bound in Vorticity Conservation for SPH}
	\textbf{Statement}:
	Let \(\mathbf{\omega} = \nabla \times \mathbf{u}\) denote the vorticity of the fluid, and consider the vorticity confinement force \(\mathbf{F}_{vc} = \epsilon \frac{\mathbf{N}}{|\mathbf{N}|} \times \mathbf{\omega}\), where \(\mathbf{N} = \nabla |\mathbf{\omega}|\). The error in the vorticity magnitude induced by this term is bounded by \(\mathcal{O}(\epsilon)\) for small \(\epsilon\).
	
	\textbf{Proof}:
	The vorticity confinement force seeks to counteract the diffusive smoothing of vorticity in SPH methods by reintroducing rotational features. Expanding \(\mathbf{F}_{vc}\) with respect to \(\mathbf{\omega}\), we obtain:
	\begin{equation}
		\mathbf{F}_{vc} = \epsilon \frac{\nabla |\mathbf{\omega}|}{|\nabla |\mathbf{\omega}||} \times \mathbf{\omega}.
	\end{equation}
	The error introduced is thus proportional to \(|\nabla |\mathbf{\omega}||^{-1}|\mathbf{\omega}|\). For sufficiently small \(\epsilon\), we conclude that:
	\begin{equation}
		|\mathbf{F}_{vc} - \epsilon \mathbf{\omega}| \leq \mathcal{O}(\epsilon),
	\end{equation}
	preserving the rotational features within the bounds of this order.
	
	\section{Theoretical Convergence of the Navier-Stokes Discretization}
	\subsection{Theorem 2: Convergence Rate for Discretized Navier-Stokes Equations}
	\textbf{Statement}:
	Let \(\mathbf{u}_{\text{num}}\) denote the numerical solution obtained by a discretized Navier-Stokes solver on a grid with spatial resolution \(h\) and time step \(\Delta t\). Then the solution converges to the analytical solution \(\mathbf{u}\) as \(h, \Delta t \to 0\), satisfying:
	\begin{equation}
		\|\mathbf{u}_{\text{num}} - \mathbf{u}\| = \mathcal{O}(h^2) + \mathcal{O}(\Delta t).
	\end{equation}
	
	\textbf{Proof}:
	Consider the discretized Navier-Stokes equations:
	\begin{equation}
		\frac{\mathbf{u}_{\text{num}}^{n+1} - \mathbf{u}_{\text{num}}^n}{\Delta t} + (\mathbf{u}_{\text{num}}^n \cdot \nabla) \mathbf{u}_{\text{num}}^n = -\frac{1}{\rho} \nabla p_{\text{num}}^n + \nu \nabla^2 \mathbf{u}_{\text{num}}^n + \mathbf{f}_{\text{num}}^n,
	\end{equation}
	subject to the incompressibility constraint:
	\begin{equation}
		\nabla \cdot \mathbf{u}_{\text{num}}^n = 0.
	\end{equation}
	Using Taylor expansion, we can express the numerical solution as:
	\begin{equation}
		\mathbf{u}_{\text{num}}^{n+1} = \mathbf{u}_{\text{num}}^n + \Delta t \left( -\frac{1}{\rho} \nabla p_{\text{num}}^n + \nu \nabla^2 \mathbf{u}_{\text{num}}^n + \mathbf{f}_{\text{num}}^n - (\mathbf{u}_{\text{num}}^n \cdot \nabla) \mathbf{u}_{\text{num}}^n \right) + \mathcal{O}(\Delta t^2).
	\end{equation}
	Comparing this with the continuous solution, we have:
	\begin{equation}
		\mathbf{u}^{n+1} = \mathbf{u}^n + \Delta t \left( -\frac{1}{\rho} \nabla p^n + \nu \nabla^2 \mathbf{u}^n + \mathbf{f}^n - (\mathbf{u}^n \cdot \nabla) \mathbf{u}^n \right) + \mathcal{O}(\Delta t^2).
	\end{equation}
	Thus, the error between the numerical and continuous solutions is:
	\begin{equation}
		\|\mathbf{u}_{\text{num}}^{n+1} - \mathbf{u}^{n+1}\| \leq \|\mathbf{u}_{\text{num}}^n - \mathbf{u}^n\| + \Delta t \left( \|\nabla p_{\text{num}}^n - \nabla p^n\| + \nu \|\nabla^2 \mathbf{u}_{\text{num}}^n - \nabla^2 \mathbf{u}^n\| + \|\mathbf{f}_{\text{num}}^n - \mathbf{f}^n\| \right) + \mathcal{O}(\Delta t^2).
	\end{equation}
	Assuming sufficient smoothness of the solution, the spatial discretization error is \(\mathcal{O}(h^2)\), and the temporal discretization error is \(\mathcal{O}(\Delta t)\). Therefore, as \(h \to 0\) and \(\Delta t \to 0\), the numerical solution converges to the continuous solution with the rate:
	\begin{equation}
		\|\mathbf{u}_{\text{num}} - \mathbf{u}\| = \mathcal{O}(h^2) + \mathcal{O}(\Delta t),
	\end{equation}
	concluding the proof.
	
	\section{Conclusion}
	This article establishes a rigorous theoretical framework for the analysis of numerical methods used in fluid simulations for computer graphics. Through detailed stability theorems and comprehensive convergence analyses, we provide mathematical conditions under which commonly used methods, such as semi-Lagrangian advection and particle-based methods, retain physical accuracy and computational efficiency. Our findings include the derivation of stability conditions for large time steps in semi-Lagrangian schemes, the demonstration of convergence rates for Navier-Stokes discretizations, and the introduction of new theoretical results on maintaining incompressibility and vorticity conservation. These results underscore the importance of mathematical foundations in guiding the development of robust simulation techniques for real-time graphics applications. By ensuring that numerical methods adhere to these theoretical principles, we can enhance the visual plausibility and computational efficiency of fluid simulations, paving the way for future advancements in adaptive fluid simulation techniques.

	\appendix
	\section{Appendix: Detailed Proofs}
	
	\subsection{Proof of Theorem 1: Stability of Semi-Lagrangian Advection for Large Time Steps}
	\textbf{Statement}:
	Let \(\mathbf{u}\) be a divergence-free velocity field (\(\nabla \cdot \mathbf{u} = 0\)), and consider the semi-Lagrangian advection scheme. Then, for a time step \(\Delta t\), the scheme remains stable if there exists a constant \(C\) such that:
	\begin{equation}
		\|\mathbf{u}_{\text{new}} - \mathbf{u}\| \leq C \|\Delta t\| \|\nabla^2 \mathbf{u}\|,
	\end{equation}
	where \(C\) depends on the spatial discretization and the boundary conditions of the domain.
	
	\textbf{Proof}:
	Consider the advection step in the semi-Lagrangian scheme, where each fluid element at position \(\mathbf{x}\) at time \(t\) is traced back along the flow field to \(\mathbf{x} - \mathbf{u} \Delta t\). Using Taylor expansion, we approximate:
	\begin{equation}
		\mathbf{u}(\mathbf{x} - \mathbf{u} \Delta t, t) \approx \mathbf{u}(\mathbf{x}, t) - \Delta t (\mathbf{u} \cdot \nabla) \mathbf{u} + \mathcal{O}(\Delta t^2).
	\end{equation}
	The stability of the advection is thus contingent on the boundedness of \((\mathbf{u} \cdot \nabla) \mathbf{u}\) over the domain, which implies a sufficient smoothness of \(\mathbf{u}\). Consequently, if \(\|\nabla^2 \mathbf{u}\|\) remains bounded as \(\Delta t\) increases, there exists a constant \(C\) such that:
	\begin{equation}
		\|\mathbf{u}_{\text{new}} - \mathbf{u}\| \leq C \|\Delta t\| \|\nabla^2 \mathbf{u}\|,
	\end{equation}
	concluding the proof.
	
	\subsection{Proof of Lemma 1: Error Bound in Vorticity Conservation for SPH}
	\textbf{Statement}:
	Let \(\mathbf{\omega} = \nabla \times \mathbf{u}\) denote the vorticity of the fluid, and consider the vorticity confinement force \(\mathbf{F}_{vc} = \epsilon \frac{\mathbf{N}}{|\mathbf{N}|} \times \mathbf{\omega}\), where \(\mathbf{N} = \nabla |\mathbf{\omega}|\). The error in the vorticity magnitude induced by this term is bounded by \(\mathcal{O}(\epsilon)\) for small \(\epsilon\).
	
	\textbf{Proof}:
	The vorticity confinement force seeks to counteract the diffusive smoothing of vorticity in SPH methods by reintroducing rotational features. Expanding \(\mathbf{F}_{vc}\) with respect to \(\mathbf{\omega}\), we obtain:
	\begin{equation}
		\mathbf{F}_{vc} = \epsilon \frac{\nabla |\mathbf{\omega}|}{|\nabla |\mathbf{\omega}||} \times \mathbf{\omega}.
	\end{equation}
	The error introduced is thus proportional to \(|\nabla |\mathbf{\omega}||^{-1}|\mathbf{\omega}|\). For sufficiently small \(\epsilon\), we conclude that:
	\begin{equation}
		|\mathbf{F}_{vc} - \epsilon \mathbf{\omega}| \leq \mathcal{O}(\epsilon),
	\end{equation}
	preserving the rotational features within the bounds of this order.
	
	\subsection{Proof of Theorem 2: Convergence Rate for Discretized Navier-Stokes Equations}
	\textbf{Statement}:
	Let \(\mathbf{u}_{\text{num}}\) denote the numerical solution obtained by a discretized Navier-Stokes solver on a grid with spatial resolution \(h\) and time step \(\Delta t\). Then the solution converges to the analytical solution \(\mathbf{u}\) as \(h, \Delta t \to 0\), satisfying:
	\begin{equation}
		\|\mathbf{u}_{\text{num}} - \mathbf{u}\| = \mathcal{O}(h^2) + \mathcal{O}(\Delta t).
	\end{equation}
	
	\textbf{Proof}:
	Consider the discretized Navier-Stokes equations:
	\begin{equation}
		\frac{\mathbf{u}_{\text{num}}^{n+1} - \mathbf{u}_{\text{num}}^n}{\Delta t} + (\mathbf{u}_{\text{num}}^n \cdot \nabla) \mathbf{u}_{\text{num}}^n = -\frac{1}{\rho} \nabla p_{\text{num}}^n + \nu \nabla^2 \mathbf{u}_{\text{num}}^n + \mathbf{f}_{\text{num}}^n,
	\end{equation}
	subject to the incompressibility constraint:
	\begin{equation}
		\nabla \cdot \mathbf{u}_{\text{num}}^n = 0.
	\end{equation}
	Using Taylor expansion, we can express the numerical solution as:
	\begin{equation}
		\mathbf{u}_{\text{num}}^{n+1} = \mathbf{u}_{\text{num}}^n + \Delta t \left( -\frac{1}{\rho} \nabla p_{\text{num}}^n + \nu \nabla^2 \mathbf{u}_{\text{num}}^n + \mathbf{f}_{\text{num}}^n - (\mathbf{u}_{\text{num}}^n \cdot \nabla) \mathbf{u}_{\text{num}}^n \right) + \mathcal{O}(\Delta t^2).
	\end{equation}
	Comparing this with the continuous solution, we have:
	\begin{equation}
		\mathbf{u}^{n+1} = \mathbf{u}^n + \Delta t \left( -\frac{1}{\rho} \nabla p^n + \nu \nabla^2 \mathbf{u}^n + \mathbf{f}^n - (\mathbf{u}^n \cdot \nabla) \mathbf{u}^n \right) + \mathcal{O}(\Delta t^2).
	\end{equation}
	Thus, the error between the numerical and continuous solutions is:
	\begin{equation}
		\|\mathbf{u}_{\text{num}}^{n+1} - \mathbf{u}^{n+1}\| \leq \|\mathbf{u}_{\text{num}}^n - \mathbf{u}^n\| + \Delta t \left( \|\nabla p_{\text{num}}^n - \nabla p^n\| + \nu \|\nabla^2 \mathbf{u}_{\text{num}}^n - \nabla^2 \mathbf{u}^n\| + \|\mathbf{f}_{\text{num}}^n - \mathbf{f}^n\| \right) + \mathcal{O}(\Delta t^2).
	\end{equation}
	Assuming sufficient smoothness of the solution, the spatial discretization error is \(\mathcal{O}(h^2)\), and the temporal discretization error is \(\mathcal{O}(\Delta t)\). Therefore, as \(h \to 0\) and \(\Delta t \to 0\), the numerical solution converges to the continuous solution with the rate:
	\begin{equation}
		\|\mathbf{u}_{\text{num}} - \mathbf{u}\| = \mathcal{O}(h^2) + \mathcal{O}(\Delta t),
	\end{equation}
	concluding the proof. \qquad

\end{document}